\documentclass[11pt,a4paper,final]{article}
\usepackage{amsmath}%
\usepackage{amstext}%
\usepackage{amssymb}%
\usepackage{showkeys}%
\usepackage{epsfig}%
\usepackage{graphicx}%
\usepackage{multicol}
\usepackage{xcolor}
\usepackage{apacite}
\usepackage{amsthm}
\usepackage{actuarialangle}
\usepackage{lifecon}
\usepackage{url}
\usepackage[top=1in, bottom=1.25in, left=0.8in, right=0.8in]{geometry}

\theoremstyle{prop}
\theoremstyle{proof}

\begin{document}

\title{Merton's Default Risk Model for Private Company}
\author{Battulga Gankhuu\footnote{
Department of Applied Mathematics, National University of Mongolia, Ulaanbaatar, Mongolia;
E-mail: battulga.g@seas.num.edu.mn}}
\date{}


\maketitle

\begin{abstract}
Because the asset value of a private company does not observable except in quarterly reports, the structural model has not been developed for a private company. For this reason, this paper attempt to develop the Merton's structural model for the private company by using the dividend discount model (DDM). In this paper, we obtain closed--form formulas of risk--neutral equity and liability values and default probability for the private company. Also, the paper provides ML estimators and the EM algorithm of our model's parameters.\\[3ex]


\end{abstract}

\section{Introduction}

Dividend discount models (DDMs), first introduced by \citeA{Williams38}, are common methods for equity valuation. The basic idea is that the market value of an equity of a firm is equal to the present value of a sum of dividend paid by the firm and price of the firm, which correspond to the next period. The same idea can be used to value the liabilities of the firm. As the outcome of DDMs depends crucially on dividend payment forecasts, most research in the last few decades has been around the proper estimations of dividend development. Also, parameter estimation of DDMs is a challenging task. Recently, \citeA{Battulga22a} introduced parameter estimation methods for practically popular DDMs. An interesting review of some existing DDMs that include deterministic and stochastic models can be found in \citeA{dAmico20}. 

Existing stochastic DDMs have one common disadvantage: If dividend and debt payments have chances to take negative values, then the market values the firm's equity and liability can take negative values with a positive probability, which is the undesirable property for the market values. A log version of the stochastic DDM, which is called by dynamic Gordon growth model was introduced by \citeA{Campbell88}, who derived a connection between log price, log dividend, and log return by approximation. Since their model is in a log framework, the stock price and dividend get positive values. For this reason, by augmenting the dynamic Gordon growth model, \citeA{Battulga22e} obtained pricing and hedging formulas, which depend on economic variables of some options and equity--linked life insurance products for a public company. For a private company, using the log private company valuation model, based on the dynamic Gordon growth model, \citeA{Battulga22d} developed closed--form pricing and hedging formulas for the European options and equity--linked life insurance products and valuation formula.

The Kalman filtering, which was introduced by \citeA{Kalman60} is an algorithm that provides estimates of some observed and unobserved (state) processes. In econometrics, the state--space model is defined by measurement and transition equations. The Kalman filtering can be used to estimate parameters and make inferences (filtering, smoothing and forecasting) about the state--space model, see \citeA{Hamilton94} and \citeA{Lutkepohl05}. Recently, to estimate parameters of a private company, \citeA{Battulga22d,Battulga22c} applied the Kalman filtering.

Default risk is a possibility that a borrower fails to make full and timely payments of principal and interest, which are stated in the debt contract. The structural model of default risk relates to option pricing. In this model, a default threshold, which represents the liabilities of the company is seen as a strike price and a asset value of the company is seen as underlying asset of the European option. For this reason, this approach is also referred to as the firm--value approach or the option--theoretic approach. Original idea of the structural model goes back to \citeA{Black73} and \citeA{Merton74}. \citeA{Black73} developed a closed--form formula for evaluating the European option and \citeA{Merton74} obtained pricing formula for the liabilities of a company under Black--Scholes framework.

In Section 2 of the paper, we develop stochastic DDM, which is known as the log private company valuation model for market values of equity and liability of a company using the \citeauthor{Campbell88}'s \citeyear{Campbell88} approximation method. Then, we model the market value of asset of the company using the approximation method once again. In Section 3, we obtain closed--form pricing formulas of the European call and put options on the market value of the asset. After that, we develop formulas of risk--neutral equity and debt values, and default probability for a private company. To the best of our knowledge, the formulas of the default risk have not been explored before. In Section 4, we study ML estimators and the EM algorithm, which are based on the Kalman filtering of our model's parameters. In Section 5, we conclude the study.

\section{Market Value Model of Equity and Liability}

Let $(\Omega,\mathcal{G}_T,\mathbb{P})$ be a complete probability space, where $\mathbb{P}$ is a given physical or real--world probability measure and $\mathcal{G}_T$ will be defined below. Dividend discount models (DDMs), first introduced by \citeA{Williams38}, are a popular tool for equity valuation. The basic idea of all DDMs is that the market value of equity at time $t-1$ of the firm equals the sum of the market value of equity at time $t$ and dividend payment at time $t$ discounted at risk--adjusted rates (required rate of return on stock). Therefore, for successive market values of equity of the company, the following relation holds 
\begin{equation}\label{eq001}
V_t^e=(1+k^e)V_{t-1}^e-p_t^e,~~~t=1,\dots,T,
\end{equation}
where $k^e$ is the required rate of return on the equity value (investors), $V_t^e$ is the market value of equity, and $p_t^e$ is the dividend payment for investors, respectively, at time $t$ of the company. On the other hand, to model market values of liabilities of the company, it is the well known fact that successive values of a debt of company or individual is given by the following equation
\begin{equation}\label{ad001}
D_t=(1+i)D_{t-1}-d_t
\end{equation}
where $D_t$ is a debt value at time $t$, $d_t$ is a debt payment at time $t$, and $i$ is a interest rate of the debt, see, e.g., \citeA{Gerber97}. Note that $D_t$ represents the principal outstanding, that is, the remaining debt immediately after $r_t$ has been paid and debt equation \eqref{ad001} shares same formula with market value of equity given in equation \eqref{eq001}. The idea of equation \eqref{ad001} can be used to model a value of liabilities of the company, namely,
\begin{equation}\label{ad002}
V_t^\ell=(1+k^\ell)V_{t-1}^\ell-p^\ell, ~~~t=1,\dots,T,
\end{equation}
where $k^\ell$ is the required rate of return on the liability value (debtholders), $V_t^\ell$ is the market value of the liability, and $p_t^\ell$ is a debt payment, which includes interest payment for debt holders, respectively, at time $t$ of the company. Note that following the idea in \citeA{Battulga22c} one can model the required rate of returns of investors and debtholders by a linear equation, which depends on economic variables. 

If payments of dividend and debt have chances to take negative values, then the market values of equity and liability of the company can take negative values with a positive probability, which is the undesirable property for market values of equity and liability. That is why, we follow the idea in \citeA{Campbell88}. As a result, the market values of equity and liability of the company take positive values. Following the idea in \citeA{Campbell88} (see also \citeA{Battulga22c}), one can obtain the following approximation
\begin{equation}\label{eq004}
\exp\{\tilde{k}\}=(V_t+p_t)\oslash V_{t-1}\approx \exp\Big\{\tilde{V}_t-\tilde{V}_{t-1}+\ln(g_t)+G_t^{-1}(G_t-I_2)\big(\tilde{p}_t-\tilde{V}_t-\mu_t\big)\Big\},
\end{equation}
where $\oslash$ is a component--wise division of two vectors, $I_n$ is an $(n\times n)$ identity matrix, $\tilde{k}:=\big(\ln(1+k^e),\ln(1+k^\ell)\big)'$ is a ($2\times 1$) log required rate of return vector, $V_t:=\big(V_t^e,V_t^\ell\big)'$ is a $(2\times 1)$ market value process at time $t$, $p_t:=\big(p_t^e,p_t^\ell\big)'$ is a $(2\times 1)$ payment process at time $t$, $\tilde{V}_t:=\ln(V_t)$ is a $(2\times 1)$ log market value process at time $t$, $\tilde{p}_t:=\ln(p_t)$ is a $(2\times 1)$ log payment process at time $t$, $\mu_t:=\mathbb{E}\big[\tilde{p}_t-\tilde{V}_t\big|\mathcal{F}_0\big]$ is a $(2\times 1)$ mean log payment--to--market value process at time $t$ of the company and $\mathcal{F}_0$ is initial information, which will be defined below, $g_t:=i_2+\exp\{\mu_t\}$ is a $(2\times 1)$ linearization parameter with $i_n=(1,\dots,1)'$ is an $(n\times 1)$ vector, whose elements equal 1, and $G_t:=\text{diag}\{g_t\}$ is a $(2\times 2)$ diagonal matrix, whose diagonal elements are $g_t$. As a result, for the log market value process at time $t$, the following approximation holds
\begin{equation}\label{eq005}
\tilde{V}_t\approx G_t(\tilde{V}_{t-1}-\tilde{p}_t+\tilde{k})+\tilde{p}_t-h_t.
\end{equation}
where $h_t:=G_t\big(\ln(g_t)-\mu_t\big)+\mu_t$ is a linearization parameter, and the model is called by dynamic Gordon growth model, see \citeA{Campbell88}. To estimate parameters of the dynamic Gordon growth model and to price the Black--Scholes call and put options on an asset value of the company, we must add a random component, namely, $u_t$, to equation \eqref{eq005}. In this case, equation \eqref{eq005} becomes 
\begin{equation}\label{eq006}
\tilde{V}_t=G_t(\tilde{V}_{t-1}-\tilde{p}_t+\tilde{k})+\tilde{p}_t-h_t+u_t.
\end{equation}

Let $B_t^e$ be a book value of equity, $B_t^\ell$ be a book value of the liability, $b_t^e$ be a book value of equity growth rate, and $b_t^\ell$ be a book value of liability growth rate, respectively, at time $t$ of the company. Since the book values of equity and liability at time $t-1$ grows at rates $b_t^e$ and $b_t^\ell$, respectively, their log values at time $t$ lead to the following vector relationship
\begin{equation}\label{eq007}
\ln(B_t)=\tilde{b}_t+\ln(B_{t-1}),
\end{equation}
where $B_t:=\big(B_t^e,B_t^\ell\big)'$ is a $(2\times 1)$ book value process and $\tilde{b}_t:=\big(\ln(1+b_t^e),\ln(1+b_t^e)\big)'$ is a $(2\times 1)$ log book value growth rate process, respectively, at time $t$ of the company. On the other side, according to \citeA{Battulga22d}, the payment process is modeled by  
\begin{equation}\label{eq008}
\tilde{p}_t=\tilde{\varrho}_t+\ln(B_{t-1}),
\end{equation}
where $\tilde{\varrho}_t:=\tilde{p}_t-\ln(B_{t-1})$ is a log payment--to--book value process at time $t$. In this paper, to error random vectors of the transition and measurement equations of the Kalman filtering, which will appear in Section 4 are independent, we assume that values of the log payment--to--book value process at times $1,\dots,T$ are known at time zero.  

Let $\tilde{m}_t:=\ln(V_t\oslash B_t)$ be a log market value--to--book value process at time $t$. Henceforth, we refer to the log market value--to--book value process $\tilde{m}_t$ as a log multiplier process. Note that for the private company, values of the log multiplier process, which is known as the state process in the Kalman filtering are unobserved. We assume that the log multiplier of the company follows the unit--root process with drift, that is, $\tilde{m}_t=\phi+\tilde{m}_{t-1}+v_t$, see \citeA{Battulga22d}. If we substitute equations $\tilde{V}_t=\tilde{m}_t+\tilde{b}_t+\ln(B_{t-1})$, $\tilde{V}_{t-1}=\tilde{m}_{t-1}+\ln(B_{t-1})$, and $\tilde{p}_t=\tilde{\varrho}_t+\ln(B_{t-1})$ into equation \eqref{eq006}, then we obtain the log private company valuation model
\begin{equation}\label{eq009}
\begin{cases}
\tilde{b}_t=-\tilde{m}_t+G_t\tilde{m}_{t-1}+c_t+u_t\\
\tilde{m}_t=\phi+\tilde{m}_{t-1}+v_t
\end{cases}~~~\text{for}~t=1,\dots,T
\end{equation}
under the real probability measure $\mathbb{P}$, where $c_t:=G_t\tilde{k}-(G_t-I_2)\tilde{\varrho}_t-h_t$ is a deterministic process. For the log private company model, where sometimes the payments are paid and sometimes not paid, we refer to \citeA{Battulga22d}. Theoretically, one can augment model \eqref{eq009} by adding an equation, which depends on economic variables including risk--free rate, see \citeA{Battulga22e} and \citeA{Battulga22b}.

Finally, let us model the market value of the asset of the company. Since the market value of the asset equals a sum of the market values of equity and liability, we have 
$$V_t^a=V_t^e+V_t^\ell,$$
where $V_t^a$ is the market value of asset of the company at time $t$. Using the same approximation method, a log asset value process of the company is approximated by the following equation
\begin{equation}\label{ad002}
\tilde{V}_t^a:=\ln(V_t^e+V_t^\ell)\approx w_t^a\tilde{V}_t^e+(1-w_t^a)\tilde{V}_t^\ell+w_t^ah_t^a 
\end{equation}
where $\mu_t^a:=\mathbb{E}[\tilde{V}_t^e-\tilde{V}_t^\ell|\mathcal{F}_0]$ is a mean log equity value--to--liability value ratio, $g_t^a:=1+\exp\{\mu_t^a\}$ and $h_t^a:=g_t^a(\ln(g_t^a)-\mu_t^a)+\mu_t^a$ are linearization parameters for log asset process, and $w_t^a=1/g_t^a$ is a weight of the approximation, respectively, at time $t$ of the company.

The stochastic properties of systems \eqref{eq009} and \eqref{ad002} are governed by the random variables $\{u_1,\dots,u_T$, $v_1,\dots,v_T,\tilde{m}_0\}$. Throughout the paper, we assume that the error random vectors $u_t$ and $v_t$ for $t=1,\dots,T$ and the initial log multiplier vector $\tilde{m}_0$ are mutually independent, and follow a normal distribution, namely,
\begin{equation}\label{eq012}
\tilde{m}_0\sim \mathcal{N}(\mu_0,\Sigma_0), ~~~u_t\sim \mathcal{N}(0,\Sigma_u),~~~v_t\sim \mathcal{N}(0,\Sigma_v) ~~~\text{for}~t=1,\dots,T
\end{equation}
under the real probability measure $\mathbb{P}$.

\section{Merton's Structural Model}

The Merton's model \citeyear{Merton74} is one of the structural models used to measure credit risk. \citeA{Merton74} was aim to value the liabilities of a specific company. As mentioned above the model connects the European call and put options.  The European call and put options are contracts that give their owner the right, but not the obligation, to buy or sell shares of a stock of a company at a predetermined price by a specified date. Let us start this Section by considering the valuation method of the European options on the asset value of a company.

Let $T$ be a time to maturity of the European call and put options, and for $t=1,\dots,T$, $\xi_t:=(u_t',v_t')'$ be a $(4\times 1)$ random error process of system \eqref{eq009}. According to equation \eqref{eq012}, $\xi_1,\dots,\xi_T$ is a random sequence of independent identically multivariate normally distributed random vectors with means of $(4\times 1)$ zero vector and covariance matrices of $(4\times 4)$ matrix $\Sigma:=\text{diag}\{\Sigma_u,\Sigma_v\}$. Therefore, a distribution of a residual random vector $\xi:=(\xi_1',\dots,\xi_T')'$ is given by 
\begin{equation}\label{eq013}
\xi \sim \mathcal{N}\big(0,I_T\otimes\Sigma\big),
\end{equation}
where $\otimes$ is the Kronecker product of two matrices.

Let $x:=(x_1',\dots,x_T')'$ be a $(4T\times 1)$ vector, which consists of all book value growth rate vectors and multiplier vectors of a company and whose $t$--th sub--vector is $x_t:=(\tilde{b}_t',\tilde{m}_t')'$. We define $\sigma$--fields, which play major roles in the paper: $\mathcal{F}_0:=\sigma(B_0,\tilde{\rho}_1,\dots,\tilde{\rho}_T)$ and for $t=1,\dots,T$, $\mathcal{F}_{t}:=\mathcal{F}_0\vee\sigma(\tilde{b}_1,\dots,\tilde{b}_t)$ and $\mathcal{G}_{t}:=\mathcal{F}_t\vee \sigma(\tilde{m}_0,\dots,\tilde{m}_t)$, where for generic $\sigma$--fields $\mathcal{O}_1$ and $\mathcal{O}_2$, $\mathcal{O}_1\vee \mathcal{O}_2$ is the minimal $\sigma$--field containing them. Note that $\sigma(B_0,\tilde{b}_1,\dots,\tilde{b}_t)=\sigma(B_0,B_1,\dots,B_t)$, the $\sigma$--field $\mathcal{F}_t$ represents available information at time $t$ for a private company, the $\sigma$--field $\mathcal{G}_t$ represents available information at time $t$ for a public company, and the $\sigma$-fields satisfy $\mathcal{F}_{t}\subset \mathcal{G}_{t}$ for $t=0,\dots,T$. Therefore, to price the Black--Scholes call and put options and obtain the Merton's formula of default probability, one has to use the information $\mathcal{G}_t$ for the public company and the information $\mathcal{F}_t$ for the private company. It follows from equation \eqref{eq013} that a joint density function of the random vector $x$ is given by
\begin{equation}\label{eq014}
f_x(x|\tilde{m}_0)=c\exp\bigg\{-\frac{1}{2}\sum_{t=1}^T
\big(Qx_t-Q_{t}x_{t-1}-q_t\big)
\Sigma^{-1}
\big(Qx_t-Q_{t}x_{t-1}-q_t\big)
\bigg\}
\end{equation}
under the real probability measure $\mathbb{P}$, where the constant is $c:=\frac{1}{(2\pi)^{2T}|\Sigma|^{T/2}}$, the coefficient matrices of the vectors $x_t$ and $x_{t-1}$ are
$$Q:=\begin{bmatrix}
I_2 & I_2\\
0 & I_2
\end{bmatrix}~~~\text{and}~~
Q_{t}:=\begin{bmatrix}
0 & G_t\\
0 & I_2 
\end{bmatrix}, ~~~\text{and}~~~
q_t:=\begin{bmatrix}
G_t\tilde{k}-h_t\\ \phi	
\end{bmatrix}.$$

To price the European call and put options, we need to change from the real probability measure to some risk--neutral measure. Let $r$ be a risk--free rate. According to \citeA{Pliska97} (see also \citeA{Bjork20}), a conditional expectation of a return process $(V_t+p_t)\oslash V_{t-1}-i_2$ must equal the risk--free rate vector $ri_2$ under some risk--neutral probability measure $\tilde{\mathbb{P}}$ and a filtration $\{\mathcal{G}_t\}_{t=0}^T$. Thus, it must hold
\begin{equation}\label{eq015}
\tilde{\mathbb{E}}\big[(V_t+p_t)\oslash V_{t-1}\big|\mathcal{G}_{t-1}\big]=(1+r)i_2
\end{equation}
for $t=1,\dots,T$, where $\tilde{\mathbb{E}}$ denotes an expectation under the risk--neutral probability measure $\mathbb{\tilde{P}}$. Using the ideas in \citeA{Battulga22e} and \citeA{Battulga22b}, one can convert the constant risk--free rate $r$ into a spot rate, which is time varying. If we substitute equation \eqref{eq006} into approximation equation \eqref{eq004}, then condition \eqref{eq015} is equivalent to the following condition
\begin{equation}\label{eq016}
\tilde{\mathbb{E}}\big[\exp\big\{G_t^{-1}u_t-(\tilde{r}i_2-\tilde{k})\big\}\big|\mathcal{G}_{t-1}\big]=i_2,
\end{equation}
where $\tilde{r}:=\ln(1+r)$ is a log risk--free rate. It should be noted that condition \eqref{eq016} corresponds only to the error random variable $u_t$. Thus, we need to impose a condition on the error random variable $v_t$ under the risk--neutral probability measure. This condition is fulfilled by $\tilde{\mathbb{E}}[\exp\{v_t\}|\mathcal{G}_{t-1}]=\hat{\theta}_t$ for $\mathcal{G}_{t-1}$ measurable any random variable $\hat{\theta}_t$. Because for any admissible choices of $\hat{\theta}_t$, condition \eqref{eq016} holds, the market is incomplete. But prices of the options are still consistent with the absence of arbitrage. In this paper, we assume that a joint distribution of the state variables $\tilde{m}_t$, $t=0,\dots,T$ is the same for the real probability measure $\mathbb{P}$ and the risk--neutral measure $\mathbb{\tilde{P}}$. Thus, we require that 
\begin{equation}\label{eq017}
\tilde{\mathbb{E}}\big[\exp\big\{v_t-1/2\mathcal{D}[\Sigma_v]\big\}\big|\mathcal{G}_{t-1}\big]=i_2.
\end{equation}
If we combine conditions \eqref{eq016} and \eqref{eq017}, then we have 
\begin{equation}\label{eq018}
\tilde{\mathbb{E}}\big[\exp\big\{R_t(\xi_t-\theta_t)\big\}\big|\mathcal{G}_{t-1}\big]=i_4,
\end{equation}
where $R_t:=\text{diag}\{G_t^{-1},I_2\}$ is a ($4\times 4$) diagonal matrix and $\theta_t:=\big((\tilde{G}_t(\tilde{r}i_2-\tilde{k}))',\frac{1}{2}\mathcal{D}[\Sigma_v]'\big)'$ is a $(4\times 1)$ deterministic Girsanov kernel process. To obtain the risk--neutral probability measure, we define the following state price density process:
\begin{eqnarray}\label{eq019}
L_t~|~\tilde{m}_0&:=&\prod_{m=1}^t \exp\bigg\{(\theta_{m}-\alpha_m)'\Sigma^{-1}\xi_m-\frac{1}{2}(\theta_{m}-\alpha_m)'\Sigma^{-1}(\theta_{m}-\alpha_m)\bigg\}
\end{eqnarray}
for $t=1,\dots,T,$ where $\alpha_m:=\frac{1}{2}\big((G_m^{-1}\mathcal{D}[\Sigma_u])',\mathcal{D}[\Sigma_v]'\big)'$ is a ($4\times 1$) deterministic vector. Then, $L_t$ is a martingale with respect to the filtration $\{\mathcal{G}_{t}\}_{t=0}^T$ and the real probability measure $\mathbb{P}$. Since $L_T>0$ and $\mathbb{E}[L_T|\mathcal{G}_0]=1$, we can define the following new probability measure:
\begin{eqnarray}\label{eq020}
&&\tilde{\mathbb{P}}\big(x\in B\big|\tilde{m}_0\big)=\int_B L_T(x|\tilde{m}_0)f_{x}(x|\tilde{m}_0)dx_1,\dots dx_T\nonumber\\
&&=\int_Bc\exp\bigg\{-\frac{1}{2}\sum_{t=1}^T
\big(Qx_t-Q_{t}x_{t-1}-\tilde{q}_t\big)
\Sigma^{-1}
\big(Qx_t-Q_{t}x_{t-1}-\tilde{q}_t\big)
\bigg\}dx_1,\dots dx_T
\end{eqnarray}
where $B\in \mathcal{B}(\mathbb{R}^{4T})$ is an any Borel set, $\tilde{q}_t:=q_t-\theta_t-\alpha_t$ is a ($4\times 1$) deterministic process, $f_{x}(x)$ is the joint density function of the random vector $x$ given by equation \eqref{eq014}, and $L_T$ is the state price density process at time $T$ given by equation \eqref{eq019}. Therefore, the log private company valuation model \eqref{eq009} becomes
\begin{equation}\label{eq021}
\begin{cases}
\tilde{b}_t=-\tilde{m}_t+G_t\tilde{m}_{t-1}+\tilde{c}_t+\tilde{u}_t\\
\tilde{m}_t=\phi+\tilde{m}_{t-1}+\tilde{v}_t
\end{cases}~~~\text{for}~t=1,\dots,T
\end{equation}
under the risk--neutral probability measure $\mathbb{\tilde{P}}$, where $\tilde{c}_t:=\tilde{r}G_ti_2-(G_t-I_2)\tilde{\varrho}_t-h_t-\frac{1}{2}G_t^{-1}\mathcal{D}[\Sigma_u]$ is a deterministic process, and a residual random vector $\tilde{\xi}:=(\tilde{\xi}_1',\dots,\tilde{\xi}_T')'$ with $\tilde{\xi}_t:=(\tilde{u}_t',\tilde{v}_t')'$, $t=1,\dots,T$ has the same distribution as the residual random vector $\xi$, that is, 
\begin{equation}\label{eq022}
\tilde{\xi} \sim \mathcal{N}\big(0,I_T\otimes\Sigma\big)
\end{equation}
under the risk--neutral probability measure $\mathbb{\tilde{P}}$. Comparing the two systems \eqref{eq009} and \eqref{eq021}, one can deduce that the log required rate of return changes from $\tilde{k}$ to the log risk--free rate vector $\tilde{r}i_2$, and an additional term $\frac{1}{2}G_t^{-1}\mathcal{D}[\Sigma_u]$ arises. Observe that the first line of system \eqref{eq021} is equivalent to 
\begin{equation}\label{eq002}
\tilde{V}_t=G_t(\tilde{V}_{t-1}-\tilde{p}_t+\tilde{r}i_2)+\tilde{p}_t-h_t-\frac{1}{2}G_t^{-1}\mathcal{D}[\Sigma_u]+\tilde{u}_t
\end{equation}
under the risk--neutral probability measure $\mathbb{\tilde{P}}$ and c.f. equation \eqref{eq006}.

To obtain distribution of the log asset value process at time $T$, let us rewrite system \eqref{eq021} in the following form
\begin{equation}\label{ad003}
x_t=\hat{Q}_tx_{t-1}+Q^{-1}\tilde{q}_t+Q^{-1}\tilde{\xi}_t
\end{equation}
where the matrix $\hat{Q}_t:=Q^{-1}Q_{t}$ satisfy that $\hat{Q}_{t+i}=\hat{Q}_{t+i}\hat{Q}_{t+i-1}\dots \hat{Q}_{t}$ and $\hat{Q}_{t+i}=\hat{Q}_{t+i}Q^{-1}$ for all $t=1,2,\dots$ and $i=0,1,\dots$. Therefore, by repeatedly using equation \eqref{ad003}, one gets that for $i=t+1,\dots,T$,
$$x_i=\hat{Q}_ix_t+\hat{Q}_i\tilde{q}_{t+1}+\dots+\hat{Q}_i\tilde{q}_{i-1}+Q^{-1}\tilde{q}_i+\hat{Q}_i\tilde{\xi}_{t+1}+\dots+\hat{Q}_i\tilde{\xi}_{i-1}+Q^{-1}\tilde{\xi}_i.$$
As a result, a sum of the process $x_t$ is given by
\begin{equation}\label{ad004}
\sum_{i=t+1}^Tx_i=\bigg(\sum_{i=t+1}^T\hat{Q}_i\bigg)x_t+\sum_{i=t+1}^T\bigg(Q^{-1}+\sum_{j=i+1}^T\hat{Q}_j\bigg)\tilde{q}_i+\sum_{i=t+1}^T\bigg(Q^{-1}+\sum_{j=i+1}^T\hat{Q}_j\bigg)\tilde{\xi}_i
\end{equation}
with a convention $\sum_{j=T+1}^T\hat{Q}_j=0$. To extract the random vectors $\tilde{b}_t$ and $\tilde{m}_t$ from the random vector $x_t$, let us define the following matrices: $J_b:=[I_2:0]$ is a $(2\times 4)$ matrix, whose first block is identity matrix of size 2 and other block is zero and $J_m:=[0:I_2]$ is a $(2\times 4)$ matrix, whose second block is identity matrix of size 2 and other block is zero. Then, it is clear that for $i=t+1,\dots,T$ and $j=i+1,\dots,T$, $J_b\hat{Q}_ix_t=(G_t-I_2)\tilde{m}_t$, $J_b\hat{Q}_j\tilde{q}_i=(G_j-I_2)\phi$, $J_bQ_i^{-1}\tilde{q}_i=\tilde{r}G_ii_2-h_i-\frac{1}{2}G_i^{-1}\mathcal{D}[\Sigma_u]-\phi$, $J_m\hat{Q}_ix_t=\tilde{m}_t$, $J_m\hat{Q}_j\tilde{q}_i=\phi$, and $J_mQ_i^{-1}\tilde{q}_i=\phi$. Consequently, it follows from equation \eqref{ad004} that conditional on the information $\mathcal{G}_t$, expectations of a sum of values at times $t+1,\dots,T$ of the book value growth rate process and log multiplier process at time $T$ are given by
\begin{eqnarray*}\label{•}
&&\tilde{\mathbb{E}}\big[\tilde{b}_{t+1}+\dots+\tilde{b}_T|\mathcal{G}_t\big]=\mathbb{\tilde{E}}\big[J_b(x_{t+1}+\dots+x_T)|\mathcal{G}_t\big]\nonumber\\
&&=\bigg(\sum_{i=t+1}^T(G_i-I_2)\bigg)\tilde{m}_t-\frac{(T-t)(T-t+1)}{2}\phi\\
&&+\sum_{i=t+1}^TG_i\big(\tilde{r}i_2+(i-t-1)\phi\big)-\sum_{i=t+1}^T\bigg(h_i+\frac{1}{2}G_i^{-1}\mathcal{D}[\Sigma_u]\bigg)\nonumber
\end{eqnarray*}
and
$$\mathbb{\tilde{E}}\big[\tilde{m}_T|\mathcal{G}_t\big]=\mathbb{\tilde{E}}\big[J_mx_T|\mathcal{G}_t\big]=\tilde{m}_t+(T-t)\phi.$$
As a result, since the log market value process at time $T$ can be represented by $\tilde{V}_T=\tilde{b}_{t+1}+\dots+\tilde{b}_T+\tilde{m}_T+\ln(B_t)$, its mean and covariance matrix conditional on $\mathcal{G}_t$ are given by
\begin{equation}\label{ad005}
\tilde{\mu}_{T|t}(\tilde{m}_t):=\mathbb{\tilde{E}}[\tilde{V}_T|\mathcal{G}_t]=\alpha_{T|t}\tilde{m}_t+\tilde{\beta}_{T|t}+\ln(B_t)
\end{equation}
and
\begin{eqnarray}\label{ad006}
\Sigma_{T|t}&:=&\widetilde{\text{Cov}}\big[\tilde{V}_T|\mathcal{G}_t\big]=J_mQ_T^{-1}\Sigma\bigg(\sum_{i=t+1}^T\bigg(Q_i^{-1}+\sum_{j=i+1}^T\hat{Q}_j\bigg)\bigg)'J_b'\nonumber\\
&+&\sum_{i_1=t+1}^T\sum_{i_2=t+1}^TJ_b\bigg(Q_{i_1}^{-1}+\sum_{j=i_1+1}^T\hat{Q}_j\bigg)\Sigma\bigg(Q_{i_2}^{-1}+\sum_{j=i_2+1}^T\hat{Q}_j\bigg)'J_b',
\end{eqnarray}
respectively, under the risk--neutral probability measure $\mathbb{\tilde{P}}$, where the coefficient matrix of the log multiplier process $\tilde{m}_t$ is
$$\alpha_{T|t}=\sum_{i=t+1}^TG_i-(T-t-1)I_2$$
and 
$$\tilde{\beta}_{T|t}:=\sum_{i=t+1}^TG_i\big(\tilde{r}i_2+(i-t-1)\phi\big)-\frac{(T-t-1)(T-t)}{2}\phi-\sum_{i=t+1}^T\bigg(h_i+\frac{1}{2}G_i^{-1}\mathcal{D}[\Sigma_u]\bigg).$$

According to equation \eqref{ad002}, the log asset value at time $T$ is represented by $\tilde{V}_T^a=(\bar{w}_T^a)'\tilde{V}_T+w_T^ah_T^a$, where $\bar{w}_T^a:=(w_T^a,1-w_T^a)'$ is a weight vector of the log asset value. Thus, due to equations \eqref{ad005} and \eqref{ad006}, conditional on the information $\mathcal{G}_t$, its distribution is given by
\begin{equation}\label{eq007}
\tilde{V}_T^a~|~\mathcal{G}_t\sim \mathcal{N}\big(\tilde{\mu}_{T|t}^a(\tilde{m}_t),(\sigma_{T|t}^{a})^2\big)
\end{equation}
under the risk--neutral probability measure $\tilde{\mathbb{P}}$, where $\tilde{\mu}_{T|t}^a(\tilde{m}_t):=(\bar{w}_T^a)^\prime\tilde{\mu}_{T|t}(\tilde{m}_t)+w_T^ah_T^a$ and $(\sigma_{T|t}^{a})^2:=(\bar{w}_T^a)^\prime\Sigma_{T|t}\bar{w}_T^a$ are conditional mean and variance of the log asset value $\tilde{V}_T^a$ given the information $\mathcal{G}_t$. 

Therefore, by equation \eqref{eq007} and the formulas that is used to price the Black--Scholes call and put options, see, e.g., \citeA{Battulga22d}, conditional on the information $\mathcal{G}_t$, prices at time $t$ of the Black--Sholes call and put options with maturity $T$ and strike price $L$ are given by
\begin{eqnarray}\label{eq036}
C_{T|t}(\tilde{m}_t)&=&e^{-(T-t)\tilde{r}}\mathbb{\tilde{E}}\big[\big(V_T^a-L\big)^+\big|\mathcal{G}_t\big]\nonumber\\
&=&\exp\bigg\{\tilde{\mu}_{T|t}^a(\tilde{m}_t)-(T-t)\tilde{r}+\frac{(\sigma_{T|t}^a)^2}{2}\bigg\}\Phi(d_{T|t}^1)-e^{-(T-t)\tilde{r}}L\Phi(d_{T|t}^2),
\end{eqnarray}
and
\begin{eqnarray}\label{eq037}
P_{T|t}(\tilde{m}_t)&=&e^{-(T-t)\tilde{r}}\mathbb{\tilde{E}}\big[\big(L-V_T^a\big)^+\big|\mathcal{G}_t\big]\nonumber\\
&=&e^{-(T-t)\tilde{r}}L\Phi(-d_{T|t}^2)-\exp\bigg\{\tilde{\mu}_{T|t}^a(\tilde{m}_t)-(T-t)\tilde{r}+\frac{(\sigma_{T|t}^a)^2}{2}\bigg\}\Phi(-d_{T|t}^1),
\end{eqnarray}
respectively, where $\Phi(x):=\int_{-\infty}^x\frac{1}{\sqrt{2\pi}}e^{-s^2/2}ds$ is the cumulative standard normal distribution function, $d_{T|t}^1:=\big(\tilde{\mu}_{T|t}^a(\tilde{m}_t)+(\sigma_{T|t}^a)^2-\ln(L)\big)/\sigma_{T|t}^a$ and $d_{T|t}^2:=d_{T|t}^1-\sigma_{T|t}^a$. Note that equation \eqref{eq036} and \eqref{eq037} can be used to price the call and put options for public companies because their log multiplier processes at time $t$ are known.  

To obtain prices of the call and put options, which do not depend on the log multiplier process $\tilde{m}_t$, we need distribution of the log multiplier process at time $t$ given the information $\mathcal{F}_t$. By replacing $c_t$ by $\tilde{c}_t$ in the Kalman filtering, which is given in Section 4, one obtains conditional mean and covariance matrix $\tilde{m}_{t|t}$ and $\Sigma(\tilde{m}_t|t)$ of the log multiplier process $\tilde{m}_t$ given the information $\mathcal{F}_t$, see Section 4. As a result, we have
$$\tilde{\mu}_{T|t}^a(\tilde{m}_t)~|~\mathcal{F}_t\sim \mathcal{N}\big(\tilde{\mu}_{T|t}^a,(\bar{w}_T^a)'\alpha_{T|t}\Sigma(\tilde{m}_t|t)\alpha_{T|t}'\bar{w}_T^a\big)$$ 
under the risk--neutral probability measure $\tilde{\mathbb{P}}$, where $\tilde{\mu}_{T|t}^a:=(\bar{w}_T^a)'\big((\alpha_{T|t}-I_2)\tilde{m}_{t|t}+\tilde{\beta}_{T|t}\big)+\tilde{V}_t^a+w_T^ah_T^a-w_t^ah_t^a$ is a conditional mean of the random variable $\tilde{\mu}_{T|t}^a(\tilde{m}_t)$ or equivalently, a conditional mean of the log asset value process $\tilde{V}_T^a$ given the information $\mathcal{F}_t$.  By using Lemma 1 in \citeA{Battulga22d} for the call and put option formulas, which are given by equations \eqref{eq036} and \eqref{eq037}, we obtain that for the private company, prices at time $t$ of the Black--Scholes call and put options are given by the following equations
\begin{eqnarray}\label{eq038}
&&C_{T|t}=e^{-(T-t)\tilde{r}}\mathbb{\tilde{E}}\big[\big(V_T^a-L\big)^+\big|\mathcal{F}_t\big]\nonumber\\
&&=\exp\bigg\{\tilde{\mu}_{T|t}^a-(T-t)\tilde{r}+\frac{(\tilde{\sigma}_{T|t}^a)^2}{2}\bigg\}\Phi(\tilde{d}_{T|t}^1)-e^{-(T-t)\tilde{r}}L\Phi(\tilde{d}_{T|t}^2),
\end{eqnarray}
and
\begin{eqnarray}\label{eq039}
&&P_{T|t}=e^{-(T-t)\tilde{r}}\mathbb{\tilde{E}}\big[\big(L-V_T^a\big)^+\big|\mathcal{F}_t\big]\nonumber\\
&&=e^{-(T-t)\tilde{r}}L\Phi(-\tilde{d}_{T|t}^2)-\exp\bigg\{\tilde{\mu}_{T|t}^a-(T-t)\tilde{r}+\frac{(\tilde{\sigma}_{T|t}^a)^2}{2}\bigg\}\Phi(-\tilde{d}_{T|t}^1),
\end{eqnarray}
respectively, where $(\tilde{\sigma}_{T|t}^a)^2:=(\bar{w}_T^a)'\big(\Sigma_{T|t}+\alpha_{T|t}\Sigma(\tilde{m}_t|t)\alpha_{T|t}'\big)\bar{w}_T^a$, $\tilde{d}_{T|t}^1:=\big(\tilde{\mu}_{T|t}^a+(\tilde{\sigma}_{T|t}^a)^2-\ln(L)\big)/\tilde{\sigma}_{T|t}^a$, and $\tilde{d}_{T|t}^2:=\tilde{d}_{T|t}^1-\tilde{\sigma}_{T|t}^a$.

Let us assume that the market value of asset of the company fully recovers when it bankrupts. Then, because values at time $T$ of the equity and debt are given by the following equations
$$V_T^e=\max(V_T^a-L,0)=(V_T^a-L)^+~~~\text{and}~~~L_T=\min(V_T^a,L)=L-(L-V_T^a)^+,$$
respectively, where $L$ is a nominal value of the debt at maturity $T$, according to formulas of the call and put options given in equations \eqref{eq038} and \eqref{eq039}, risk--neutral market values of the equity and debt at time $t$ are given by
$$\bar{V}_t^e=C_{T|t}~~~\text{and}~~~L_t=Le^{-(T-t)\tilde{r}}-P_{T|t}.$$
If we equate the equity value and the risk--neutral equity value, then one may obtain one possible version of estimation for the default threshold $\bar{L}$ (see \citeA{McNeil15}) from the following equation
$$\exp\{\tilde{m}_{0|0}^e\}B_0^e=C_{T|0}(\bar{L}),$$
where $\tilde{m}_{0|0}^e$ is a final smoothed equity multiplier, which is based on last $T$ estimation periods, see Section 4.

Now, we move to a default probability of a company. In order to obtain the default probability of the company, we need a distribution of log asset value at time $T$ given the information $\mathcal{G}_t$. By using the same idea as mentioned above, one gets that conditional on the information $\mathcal{G}_t$, its distribution is given by
\begin{equation}\label{ad007}
\tilde{V}_T^a~|~\mathcal{G}_t\sim \mathcal{N}\big(\mu_{T|t}^a(\tilde{m}_t),(\sigma_{T|t}^{a})^2\big)
\end{equation}
under the real probability measure $\mathbb{P}$, where the conditional expectation is given by 
$$\mu_{T|t}^a(\tilde{m}_t):=(\bar{w}_T^a)'\big((\alpha_{T|t}-I_2)\tilde{m}_t+\beta_{T|t}\big)+\tilde{V}_t^a+w_T^ah_T^a-w_t^ah_t^a$$ with
$$\beta_{T|t}:=\sum_{i=t+1}^TG_i\big(\tilde{k}+(i-t-1)\phi\big)-\frac{(T-t-1)(T-t)}{2}\phi-\sum_{i=t+1}^Th_i.$$
According to the structural model of default risk, if the asset value of a company falls below the default threshold, representing liabilities, then default occurs. Therefore, due to equation \eqref{ad007}, conditional on the information $\mathcal{G}_t$, the default probability at time $t$ of the company is given by the following equation

\begin{equation}\label{eq008}
\mathbb{P}\big[V_T^a\leq \bar{L}|\mathcal{G}_t\big]=\mathbb{P}\big[\tilde{V}_T^a\leq \ln(\bar{L})|\mathcal{G}_t\big]=\Phi\left(\frac{\ln(\bar{L})-\mu_{T|t}^a(\tilde{m}_t)}{\sigma_{T|t}^a}\right),
\end{equation}
where $\bar{L}$ is the default threshold at maturity $T$. Of course this formula can be used to calculate the default probability of a public company.

To obtain the Merton's default probability for the private company, one needs a distribution of the log multiplier process given the information $\mathcal{F}_t$ under the real probability measure $\mathbb{P}$. For the conditional mean and covariance matrix $\tilde{m}_{t|t}$ and $\Sigma(\tilde{m}_t|t)$ of the log multiplier process $\tilde{m}_t$ given the information $\mathcal{F}_t$, since the default probability is calculated in the real world, they are obtained from the Kalman filtering without replacement as compared to the option valuation. Thus, one gets that
$$\mu_{T|t}^a(\tilde{m}_t)~|~\mathcal{F}_t\sim \mathcal{N}\big(\mu_{T|t}^a,(\bar{w}_T^a)'\alpha_{T|t}\Sigma(\tilde{m}_t|t)\alpha_{T|t}'\bar{w}_T^a\big)$$ 
under the real probability measure $\tilde{\mathbb{P}}$, where $\mu_{T|t}^a:=(\bar{w}_T^a)'\big((\alpha_{T|t}-I_2)\tilde{m}_{t|t}+\beta_{T|t}\big)+\tilde{V}_t^a+w_T^ah_T^a-w_t^ah_t^a$ is a conditional mean of the random variable $\mu_{T|t}^a(\tilde{m}_t)$ given the information $\mathcal{F}_t$. Consequently, it follows from Lemma 1 in \citeA{Battulga22d} and equation \eqref{eq008} that the default probability of the private company is given by  
$$\mathbb{P}\big[V_T^a\leq \bar{L}|\mathcal{F}_t\big]=\mathbb{E}\big[\mathbb{P}\big[V_T^a\leq \bar{L}|\mathcal{G}_t\big]\big|\mathcal{F}_t\big]=\Phi\left(\frac{\ln(\bar{L})-\mu_{T|t}^a}{\tilde{\sigma}_{T|t}^a}\right).$$

\section{The Kalman Filtering}

Let us reconsider the log private company valuation model \eqref{eq009}. The model can be written by state--space model
\begin{equation}\label{eq066}
\begin{cases}
\tilde{b}_t=\Psi_t z_t+c_t+u_t\\
z_t=Az_{t-1}+a+\eta_t
\end{cases}~~~\text{for}~t=1,\dots,T,
\end{equation}
where $z_t:=(\tilde{m}_t',\tilde{m}_{t-1}')'$ is a ($4\times 1$) state process of the multipliers at times $t$ and $t-1$, $a:=(\phi,0)'$ is a ($4\times 1$) constant vector, $\eta_t:=(v_t',0)'$ is a ($4\times 1$) random error process, whose covariance matrix equals $\Sigma_\eta:=\text{diag}\{\Sigma_v,0\}$, and
$$\Psi_t:=\begin{bmatrix}
-I_2 & G_t\\
0 & 0
\end{bmatrix}~~~\text{and}~~~A:=\begin{bmatrix}
I_2 & 0\\
I_2 & 0
\end{bmatrix}
$$
are ($4\times 4$) matrices. Note that adding equation $r_t=r+w_t$, which is independent of error random vector $v_t$ into the state--space model \eqref{eq066}, one may estimate the risk--free rate. For system \eqref{eq066}, its first line determines the measurement equation and the second line determines the transition equation. For each $t=0,\dots,T$, conditional on the information $\mathcal{F}_t$, conditional expectations and covariance matrices of the log book value growth rate process and the state process are recursively obtained by the Kalman filtering (see \citeA{Hamilton94} and \citeA{Lutkepohl05}):
\begin{itemize}
\item Initialization: 
\begin{itemize}
\item Expectation
\begin{equation}\label{eq067}
z_{0|0}:=\mathbb{E}[z_0|\mathcal{F}_0]=(\mu_0',\mu_0')'
\end{equation}
\item Covariance
\begin{equation}\label{eq068}
\Sigma(z_0|0):=\text{Cov}[z_0|\mathcal{F}_0]=\text{diag}\{\Sigma_0,\Sigma_0\}
\end{equation}
\end{itemize}
\item Prediction step: for $t=1,\dots,T$, 
\begin{itemize}
\item Expectations
\begin{eqnarray}
z_{t|t-1}&:=&\mathbb{E}(z_t|\mathcal{F}_{t-1})=Az_{t-1|t-1}+a\label{eq069}\\
\tilde{b}_{t|t-1}&:=&\mathbb{E}(\tilde{b}_t|\mathcal{F}_{t-1})=\Psi_t a+c_t+\Psi_t Az_{t-1|t-1}\label{eq070}
\end{eqnarray}
\item Covariances
\begin{eqnarray}
\Sigma(z_{t}|t-1)&:=&\text{Cov}[z_{t}|\mathcal{F}_{t-1}]=A\Sigma(z_{t-1}|t-1)A'+\Sigma_\eta\label{eq071}\\
\Sigma(\tilde{b}_t|t-1)&:=&\text{Cov}[\tilde{b}_{t}|\mathcal{F}_{t-1}]=\Psi_t\Sigma(z_{t}|t-1)\Psi_t'+\Sigma_{u}\label{eq072}
\end{eqnarray}
\end{itemize}
\item Correction step: for $t=1,\dots,T$, 
\begin{itemize}
\item Expectations
\begin{equation}\label{eq073}
z_{t|t}:=\mathbb{E}[z_t|\mathcal{F}_t]=z_{t|t-1}+\mathcal{K}_{t}(\tilde{b}_t-\tilde{b}_{t|t-1})
\end{equation}
\item Covariances
\begin{equation}\label{eq074}
\Sigma(z_t|t):=\text{Cov}[z_t|\mathcal{F}_t]=\Sigma(z_t|t-1)-\mathcal{K}_{t}\Sigma(\tilde{b}_t|t-1)\mathcal{K}_t',
\end{equation}
where $\mathcal{K}_t:=\Sigma(z_t|t-1)\Psi_t'\Sigma(\tilde{b}_t|t-1)^{-1}$ is the Kalman filter gain.
\end{itemize}
\end{itemize}

For each $t=T+1,T+2,\dots$, conditional on the information $\mathcal{F}_T$, conditional expectations and covariance matrices of the log book value growth rate process and the state process are recursively obtained by (see \citeA{Hamilton94} and \citeA{Lutkepohl05}):

\begin{itemize}
\item Forecasting step: for $t=T+1,T+2,\dots$, 
\begin{itemize}
\item Expectations
\begin{eqnarray}
z_{t|T}&:=&\mathbb{E}[z_t|\mathcal{F}_T]=Az_{t-1|T}+a\label{eq075}\\
\tilde{b}_{t|T}&:=&\mathbb{E}[\tilde{b}_t|\mathcal{F}_T]=\Psi_tz_{t|T}+c_t\label{eq076}
\end{eqnarray}
\item Covariances
\begin{eqnarray}
\Sigma(z_t|T)&:=&\text{Cov}[z_t|\mathcal{F}_T]=A\Sigma(z_{t-1}|T)A'+\Sigma_\eta\label{eq077}\\
\Sigma(\tilde{b}_t|T)&:=&\text{Cov}[\tilde{b}_t|\mathcal{F}_T]=\Psi_t\Sigma(z_t|T)\Psi_t'+\Sigma_u\label{eq078}
\end{eqnarray}
\end{itemize}
\end{itemize}

The Kalman filtering, which is considered above provides an algorithm for filtering of the state process $z_t$, which is unobserved variable. To estimate parameters of our model \eqref{eq009} except the risk--free rate $r$, in addition to the Kalman filtering, we also need to make inference about the state process $z_t$ for each $t=1,\dots,T$ based on the full information $\mathcal{F}_T$, see below. Such an inference is called the smoothed estimate of the state process $z_t$. The smoothed inference of the state process can be obtained by the following Kalman smoother recursions, see \citeA{Hamilton94} and \citeA{Lutkepohl05}.

\begin{itemize}
\item Smoothing step: for $t=T-1,T-2,\dots,0$,
\begin{itemize}
\item Expectations
\begin{equation}\label{eq079}
z_{t|T}:=\mathbb{E}[z_t|\mathcal{F}_T]=z_{t|t}+\mathcal{S}_{t}\big(z_{t+1|T}-z_{t+1|t}\big)
\end{equation}
\item Covariances
\begin{equation}\label{eq080}
\Sigma(z_t|T):=\text{Cov}[z_t|\mathcal{F}_T]=\Sigma(z_t|t)-\mathcal{S}_{t}\big(\Sigma(z_{t+1}|t)-\Sigma(z_{t+1}|T)\big)\mathcal{S}_t',
\end{equation}
where $\mathcal{S}_{t}:=\Sigma(z_t|t)A'\Sigma^{-1}(z_{t+1}|t)$ is the Kalman smoother gain. Also, it can be shown that
\begin{equation}\label{eq098}
\Sigma(z_{t},z_{t+1}|T):=\text{Cov}[z_{t},z_{t+1}|\mathcal{F}_T]=\mathcal{S}_{t}\Sigma(z_{t+1}|T),
\end{equation}
see \citeA{Battulga22c}.
\end{itemize}
\end{itemize}

In the EM algorithm, one considers a joint density function of a random vector, which is composed of observed variables and state variables. In our cases, the vectors of observed variables and the state variables correspond to a vector of the log book value growth rates, $\tilde{b}:=(\tilde{b}_1',\dots,\tilde{b}_T')'$, and a vector of the log multipliers, $\tilde{m}:=(\tilde{m}_0,\dots,\tilde{m}_T)'$, respectively. Interesting usages of the EM algorithm in econometrics can be found in \citeA{Hamilton90} and \citeA{Schneider92}. Let us denote the joint density function by $f_{\tilde{b},\tilde{m}}(\tilde{b},\tilde{m})$. The EM algorithm consists of two steps. In the expectation (E) step of the EM algorithm, one has to determine a form of an expectation of log of the joint density given the full information $\mathcal{F}_T$. We denote the expectation by $\Lambda(\theta|\mathcal{F}_T)$, that is, $\Lambda(\theta|\mathcal{F}_T):=\mathbb{E}\big[\ln\big(f_{\tilde{b},\tilde{m}}(\tilde{b},\tilde{m})\big)|\mathcal{F}_T\big]$. 
Then, for our log private company valuation model \eqref{eq009}, one can show that the expectation of log of the joint density of the vectors of the log book value growth rates and the price--to--book ratios is
\begin{eqnarray}\label{eq081}
\Lambda(\theta|\mathcal{F}_T)&=&-(2T+1)\ln(2\pi)-\frac{T}{2}\ln(|\Sigma_u|)-\frac{T}{2}\ln(|\Sigma_{v}|)-\frac{1}{2}\ln(|\Sigma_0|)-\frac{1}{2}\sum_{t=1}^T\mathbb{E}\big[u_t'\Sigma_uu_t\big|\mathcal{F}_T\big]\nonumber\\
&-&\frac{1}{2}\sum_{t=1}^T\mathbb{E}\big[v_t'\Sigma_{v}^{-1}v_t\big|\mathcal{F}_T\big]-\frac{1}{2}\mathbb{E}\big[(\tilde{m}_0-\mu_0)'\Sigma_0^{-1}(\tilde{m}_0-\mu_0)\big|\mathcal{F}_T\big],
\end{eqnarray}
where recall that the error random vectors are given by $u_t=\tilde{b}_t+\tilde{m}_t-\tilde{\varrho}_t-G_t(\tilde{m}_{t-1}-\tilde{\varrho}_t+\tilde{k})+h_t$ and $v_t=\tilde{m}_t-\phi-\tilde{m}_{t-1}$, and $\theta:=\big(\tilde{k}',\tilde{\mu}_0',\tilde{\phi}',\text{vech}(\Sigma_u)',\text{vech}(\Sigma_v)',\text{vech}(\Sigma_0)'\big)'$ is a $(15\times 1)$ vector, which consists of all parameters of the model \eqref{eq009} except the risk--free rate. 

Observe that the log payment--to--market value is given by $\tilde{p}_t-\tilde{V}_t=\tilde{\varrho}_t-\tilde{b}_t-\tilde{m}_t$. Thus, due to system \eqref{eq009}, the log payment--to--market value is represented by $\tilde{p}_t-\tilde{V}_t=h_t-G_t(\tilde{m}_{t-1}-\tilde{\varrho}_t+\tilde{k}_t)-u_t.$ Therefore, as $\mu_t=\mathbb{E}[\tilde{p}_t-\tilde{V}_t|\mathcal{F}_0]$, we get that
\begin{equation}\label{eq082}
\mu_t=h_t+G_t\big(\tilde{\varrho}_t-\tilde{k}-\mathbb{E}[\tilde{m}_{t-1}|\mathcal{F}_0]\big).
\end{equation}
Consequently, because $h_t=G_t\big(\ln(g_t)-\mu_t\big)+\mu_t=g_t\odot\big(\ln(g_t)-\mu_t\big)+\mu_t$ and $\mathbb{E}[\tilde{m}_{t-1}|\mathcal{F}_0]=\mu_0+(t-1)\phi$, one obtain that
\begin{equation}\label{eq083}
g_t=i_2\oslash (i_2-\exp\{\varphi_t\}),
\end{equation}
where $\odot$ is the Hadamard's component--wise product of two vectors and $\varphi_t:=\tilde{\varrho}_t-\tilde{k}-\big(\mu_0+(t-1)\phi\big)$ is a ($2\times 1$) vector. Further, since $\mu_t=\ln(g_t-i_2)=\varphi_t+\ln(g_t)$, we get that
\begin{equation}\label{eq084}
h_t=-\Big\{\varphi_t\odot\exp\{\varphi_t\}\oslash\big(i_2-\exp\{\varphi_t\}\big)+\ln\big(i_2-\exp\{\varphi_t\}\big)\Big\}.
\end{equation}

Let us define a vector and matrices, which deal with partial derivatives of $\Lambda(\theta|\mathcal{F}_T)$ with respect to parameters $\tilde{k}$, $\mu_0$, and $\phi$:
\begin{equation}\label{•}
d_t:=G_t(G_t-I_2)\big(\tilde{m}_{t-1}-(\mu_0+(t-1)\phi)\big),
\end{equation}
and
$$C_1:=\begin{bmatrix}
1 & 0\\
0 & 0
\end{bmatrix}~~~\text{and}~~~ C_2:=\begin{bmatrix}
0 & 0\\
0 & 1
\end{bmatrix}.$$
A conditional mean of the vector $d_t$ given full information $\mathcal{F}_T$ is
\begin{equation}\label{•}
d_{t|T}:=\mathbb{E}\big[d_t\big|\mathcal{F}_T\big]=G_t(G_t-I_2)\big(\tilde{m}_{t-1|T}-(\mu_0+(t-1)\phi)\big),
\end{equation}
where $\tilde{m}_{t|T}:=Ez_{t|T}$ is a $(2\times 1)$ smoothed inference of the log multiplier process $\tilde{m}_t$ and $E:=[I_2:0]$ is a $(2\times 4)$ matrix, which used to extract first two components of the vector $z_{t|T}$. Then, it is clear that partial derivatives of $u_t$ with respect to the parameters $\tilde{k}$, $\mu_0$, and $\phi$ are given by
\begin{equation}\label{•}
\frac{\partial u_t}{\partial \tilde{k}'}=[C_1(d_t-g_t):C_2(d_t-g_t)],~~~
\frac{\partial u_t}{\partial \mu_0'}=[C_1d_t:C_2d_t],~~~\frac{\partial u_t}{\partial \phi'}=[C_1d_t:C_2d_t](t-1).
\end{equation}

As a result, partial derivatives of $\Lambda(\theta|\mathcal{F}_T)$ with respect to the parameters $\tilde{k}$, $\mu_0$, and $\phi$ are obtained by
\begin{equation}\label{ad012}
\frac{\partial \Lambda(\theta|\mathcal{F}_T)}{\partial \tilde{k}'}=-\sum_{t=1}^T\mathbb{E}\Big\{\big[u_t'\Sigma_u^1(d_t-g_t):u_t'\Sigma_u^2(d_t-g_t)\big]\Big|\mathcal{F}_T\Big\},
\end{equation}
\begin{equation}\label{ad013}
\frac{\partial \Lambda(\theta|\mathcal{F}_T)}{\partial \mu_0'}=-\sum_{t=1}^T\mathbb{E}\Big\{\big[u_t'\Sigma_u^1d_t:u_t'\Sigma_u^2d_t\big]\Big|\mathcal{F}_T\Big\}+(\tilde{m}_{0|T}-\mu_0)'\Sigma_0^{-1},
\end{equation}
and
\begin{equation}\label{ad014}
\frac{\partial \Lambda(\theta|\mathcal{F}_T)}{\partial \phi'}=-\sum_{t=1}^T(t-1)\mathbb{E}\Big\{\big[u_t'\Sigma_u^1d_t:u_t'\Sigma_u^2d_t\big]\Big|\mathcal{F}_T\Big\}+\sum_{t=1}^Tv_{t|T}'\Sigma_v^{-1},
\end{equation}
where $\Sigma_u^i:=\Sigma_uC_i$ for $i=1,2$ and $v_{t|T}:=\tilde{m}_{t|T}-\phi-\tilde{m}_{t-1|T}$ is a smoothed residual process, corresponding to the residual process $v_t$.

For the full information $\mathcal{F}_T$, let us denote smoothed residual process, corresponding to the residual process $u_t$ by $u_{t|T}=\tilde{b}_t+\tilde{m}_{t|T}-\tilde{\varrho}_t-G_t(\tilde{m}_{t-1|T}-\tilde{\varrho}_t+\tilde{k})+h_t$ and conditional expectations of products of the state variables by $\tilde{m}_{t-1,t-1|T}:=\mathbb{E}[\tilde{m}_{t-1}\tilde{m}_{t-1}|\mathcal{F}_T]=E\big(\Sigma(z_{t-1}|T)+z_{t-1|T}z_{t-1|T}'\big)E'$ and $\tilde{m}_{t-1,t|T}:=\mathbb{E}[\tilde{m}_{t-1}\tilde{m}_{t}|\mathcal{F}_T]=E\big(\mathcal{S}_{t-1}\Sigma(z_{t}|T)+z_{t-1|T}z_{t|T}'\big)E'$, see equation \eqref{eq098}. Then, as $d_t-d_{t|T}=G_t(G_t-I_2)(\tilde{m}_{t-1}-\tilde{m}_{t-1|T})$, $u_t-u_{t|T}=\tilde{m}_t-\tilde{m}_{t|T}-G_t(\tilde{m}_{t-1}-\tilde{m}_{t-1|T})$, $\text{Cov}[\tilde{m}_{t-1},\tilde{m}_t|\mathcal{F}_T]=\tilde{m}_{t-1,t|T}+\tilde{m}_{t-1|T}\tilde{m}_{t|T}'$, and $\text{Cov}[\tilde{m}_{t-1},\tilde{m}_{t-1}|\mathcal{F}_T]=\tilde{m}_{t-1,t-1|T}+\tilde{m}_{t-1|T}\tilde{m}_{t-1|T}'$, the following equation holds
\begin{equation}\label{ad008}
\mathbb{E}\big[d_tu_t'|\mathcal{F}_T\big]=Z_t+d_{t|T}u_{t|T}'
\end{equation}
where $Z_t:=G_t(G_t-I_2)\big(\tilde{m}_{t-1,t|T}+\tilde{m}_{t-1|T}\tilde{m}_{t|T}'-(\tilde{m}_{t-1,t-1|T}+\tilde{m}_{t-1|T}\tilde{m}_{t-1|T}')G_t\big)$ is a $(2\times 2)$ matrix. From equation \eqref{ad008}, it also holds
\begin{equation}\label{ad009}
\mathbb{E}\big[(d_t-g_t)u_t'|\mathcal{F}_T\big]=Z_t+(d_{t|T}-g_t)(u_{t|T}^k)'+(d_{t|T}-g_t)\tilde{k}'G_t,
\end{equation}
where $u_{t|T}^k=\tilde{b}_t+\tilde{m}_t-\tilde{\varrho}_t-G_t(\tilde{m}_{t-1}-\tilde{\varrho}_t)+h_t$ is a smoothed process, which excludes the term $G_t\tilde{k}$ from the smoothed residual process $u_{t|T}$. Therefore, according to equations \eqref{ad008} and \eqref{ad009}, we get that
\begin{equation}\label{ad010}
\mathbb{E}\big[u_t'\Sigma_u^id_t|\mathcal{F}_T\big]=\text{tr}\big\{\Sigma_u^i\big(Z_t+d_{t|T}u_{t|T}'\big)\big\}
\end{equation}
and
\begin{equation}\label{ad011}
\mathbb{E}\big[u_t'\Sigma_u^i(d_t-g_t)|\mathcal{F}_T\big]=\text{tr}\big\{\Sigma_u^i\big(Z_t+(d_{t|T}-g_t)(u_{t|T}^k)'\big)\big\}+(d_{t|T}-g_t)'\Sigma_u^iG_t\tilde{k}
\end{equation}
for $i=1,2$, where $\text{tr}(A)$ is the trace of a square matrix $A$. If we equate equations \eqref{ad012}--\eqref{ad014} to zero, then due to equations \eqref{ad010} and \eqref{ad011}, one obtains estimators of the parameters $\tilde{k}$, $\mu_0$, and $\phi$
\begin{equation}\label{ad016}
\hat{\tilde{k}}:=\bigg(\sum_{t=1}^T\begin{bmatrix}
w_t'\Sigma_u^1(d_{t|T}-g_t)\\
w_t'\Sigma_u^2(d_{t|T}-g_t)
\end{bmatrix}
\bigg)^{-1}\sum_{t=1}^T\begin{bmatrix}
\text{tr}\big\{\Sigma_u^1\big(Z_t+(d_{t|T}-g_t)(u_{t|T}^k)'\big)\big\}\\
\text{tr}\big\{\Sigma_u^2\big(Z_t+(d_{t|T}-g_t)(u_{t|T}^k)'\big)\big\}
\end{bmatrix},
\end{equation}
\begin{equation}\label{ad017}
\hat{\mu}_0:=\tilde{m}_{0|T}-\Sigma_0\sum_{t=1}^T\begin{bmatrix}
\text{tr}\big\{\Sigma_u^1\big(Z_t+d_{t|T}u_{t|T}'\big)\big\}\\
\text{tr}\big\{\Sigma_u^2\big(Z_t+d_{t|T}u_{t|T}'\big)\big\}
\end{bmatrix},
\end{equation}
and
\begin{equation}\label{ad018}
\hat{\phi}:=\frac{1}{T}\left\{\tilde{m}_{T|T}-\tilde{m}_{0|T}-\Sigma_v\sum_{t=1}^T(t-1)\begin{bmatrix}
\text{tr}\big\{\Sigma_u^1\big(Z_t+d_{t|T}u_{t|T}'\big)\big\}\\
\text{tr}\big\{\Sigma_u^2\big(Z_t+d_{t|T}u_{t|T}'\big)\big\}
\end{bmatrix}\right\}.
\end{equation}
For estimators of the covariance matrices $\Sigma_u$, $\Sigma_v$, and $\Sigma_0$, the following formulas holds
\begin{equation}\label{ad015}
\hat{\Sigma}_u:=\frac{1}{T}\sum_{t=1}^T\mathbb{E}[u_tu_t'|\mathcal{F}_T],
~~~\hat{\Sigma}_v:=\frac{1}{T}\sum_{t=1}^T\mathbb{E}[v_tv_t'|\mathcal{F}_T], 
~~~\hat{\Sigma}_0:=\Sigma(\tilde{m}_0|T).
\end{equation}

To calculate the conditional expectations $\mathbb{E}\big(u_tu_t'|\mathcal{F}_T\big)$ and $\mathbb{E}\big(v_tv_t'|\mathcal{F}_T\big)$, observe that the random error processes at time $t$ of the log book value growth rate process and the log multiplier process can be represented by
\begin{eqnarray}\label{eq091}
u_t&=&u_{t|T}-\Psi_t(z_t-z_{t|T})\nonumber\\
v_t&=&v_{t|T}+E(z_t-z_{t|T})-EA(z_{t-1}-z_{t-1|T}).
\end{eqnarray}
Therefore, as $u_{t|T}$ and $v_{t|T}$ are measurable with respect to the full information $\mathcal{F}_T$ (known at time $T$), it follows from equations \eqref{eq098} that
\begin{eqnarray}\label{eq092}
\mathbb{E}\big(u_tu_t'|\mathcal{F}_T\big)&=&u_{t|T}u_{t|T}'+\Psi_t\Sigma(z_t|T)\Psi_t'\nonumber\\
\mathbb{E}\big(v_tv_t'|\mathcal{F}_T\big)&=&v_{t|T}v_{t|T}'+E\Sigma(z_t|T)E'+EA\Sigma(z_{t-1}|T)A'E'-2EA\mathcal{S}_{t-1}\Sigma(z_{t}|T)E',
\end{eqnarray}
c.f. \citeA{Schneider92}. If we substitute equation \eqref{eq092} into \eqref{ad015}, then under suitable conditions the zig--zag iteration that corresponds to equations \eqref{eq067}--\eqref{eq074}, \eqref{eq079}, \eqref{eq080}, and \eqref{ad016}--\eqref{ad015} converges to the maximum likelihood estimators of our log private company valuation model. As a result, an smoothed inference of the market value vector at time $t$ of the private company is calculated by the following formula
\begin{equation}\label{eq099}
V_{t|T}=m_{t|T}B_t, ~~~t=0,1,\dots,T,
\end{equation}
where $m_{t|T}=\exp\{\tilde{m}_{t|T}\}$ is a smoothed multiplier vector at time $t$. Also, an analyst can forecast the market value process of the private company by using equations \eqref{eq075} and \eqref{eq076}. 

\section{Conclusion}

Because the asset value of a private company does not observable except in quarterly reports, the structural model has not been developed for a private company. For this reason, this paper is dedicated to develop the Merton's structural model for the private company. To obtain a distribution of the market value of the asset of the company we develop the log private company valuation model for market values of equity and liability of a company using the \citeauthor{Campbell88}'s \citeyear{Campbell88} approximation method. Using the distribution of the market value of the asset, we obtain closed--form formulas of risk--neutral equity and liability values and default probability for the private company. Finally, the paper provides ML estimators and the EM algorithm of our model's parameters. The suggested model can be used not only by private companies but also by public companies

Further extensions of the model are as follows: (i) the log private company valuation model should be connected to other advanced structural models, (ii) the model should be extended to correlated multiple companies, (iii) the error term should be modeled by correlated conditional heteroscedastic models, (iv) the multiplier process should be modeled by more advanced process, e.g., AR$(p)$ process with unit root, and (v) the risk--free rate should be modeled by a model that varies over time, see \citeA{Battulga22e} and \citeA{Battulga22b}.

\bibliographystyle{apacite}
\bibliography{References}

\end{document}